\documentclass[twocolumn,floatfix]{aastex631}

\usepackage{xcolor}
\shorttitle{New recipes for measuring dynamical state}
\shortauthors{Kim et al.}
\graphicspath{{./}{figures/}}
\begin{document}

\title{New observational recipes for measuring dynamical state of galaxy clusters}

\author[0000-0003-4032-8572]{Hyowon Kim}
\affiliation{Korea Astronomy and Space science Institute, Daejeon, 34055, Korea}
\affiliation{Korea University of Science and Technology, Daejeon, 34113, Korea}

\author[0000-0001-5303-6830]{Rory Smith}
\affiliation{Departamento de F$\acute{i}$sica, Universidad T$\acute{e}$cnica Federico Santa Mar$\acute{i}$a, Santiago, Chile}

\author[0000-0002-9434-5936]{Jongwan Ko}
\affiliation{Korea Astronomy and Space science Institute, Daejeon, 34055, Korea}
\affiliation{Korea University of Science and Technology, Daejeon, 34113, Korea}

\author[0000-0001-7967-6473]{Jong-Ho Shinn}
\affiliation{Korea Astronomy and Space science Institute, Daejeon, 34055, Korea}

\author[0000-0001-9544-7021]{Kyungwon Chun}
\affiliation{Korea Astronomy and Space science Institute, Daejeon, 34055, Korea}

\author[0000-0001-5135-1693]{Jihye Shin}
\affiliation{Korea Astronomy and Space science Institute, Daejeon, 34055, Korea}

\author[0000-0002-6841-8329]{Jaewon Yoo}
\affiliation{Quantum Universe Center, Korea Institute of Advanced Science, Seoul, Korea}



\begin{abstract}
{During cluster assembly, a cluster's virialization process leaves behind signatures that can provide information on its dynamical state. However, no clear consensus yet exists on the best way to achieve this. Therefore, we attempt to derive improved recipes for classifying the dynamical state of clusters in observations using cosmological simulations. Cluster halo mass and their subhalos' mass are used to $ 10^{14}M_{\odot} h^{-1}$ and $10^{10}M_{\odot} h^{-1}$ to calculate five independent dynamical state indicators. We experiment with recipes by combining two to four indicators for detecting specific merger stages like recent and ancient mergers. These recipes are made by plotting merging clusters and a control sample of relaxed clusters in multiple indicators parameter space, and then applying a rotation matrix method to derive the best way to separate mergers from the control sample. The success of the recipe is quantified using the success rate and the overlap percentage of the merger and control histograms along the newly rotated $x$-axis. This provides us with recipes using different numbers of combined indicators and for different merger stage. Among the recipes, the stellar mass gap and center offset are the first and second most dominant of the indicators, and using more indicators improves the effectiveness of the recipe. When applied to observations, our results show good agreement with literature values of cluster dynamical state.}
\end{abstract}



\section{Introduction} \label{sec:intro}
In the hierarchical formation scenario, massive structures such as galaxy clusters are the most likely structures to be actively assembling today, and thus can be found in a wide range of dynamical states. The dynamical states of galaxy clusters can be broadly separated into two categories; relaxed (or virialized) and unrelaxed (or non-virialized). 

In the relaxed state, a cluster typically has one dominant Brightest Cluster Galaxy (BCG) located at the bottom of the deepest part of the potential well, an aligned and ordered distribution of hot gas around the BCG, and a single Gaussian velocity dispersion distribution of member galaxies \citep{2008ApJS..174..117M, 2020MNRAS.497.5485Y}. On the other hand, unrelaxed clusters frequently show two or more BCGs, a disordered hot gas distribution, multiple peaks in the velocity dispersion distribution of member galaxies, or radio features such as radio relics and radio halos \citep{2019ApJ...882...69G, 2018MNRAS.478.5473L}. Thus, all these features could be used to constrain the assembly history of a galaxy cluster.

Many previous studies used the properties of cluster galaxies to ascertain the dynamical state. For example, some previous studies used the BCG \citep{2017ApJ...836..105K, 2018MNRAS.478.5473L, 2019ApJ...887..264R, 2021MNRAS.504.5383D}, redshift distribution \citep{1988AJ.....95..985D, 2021MNRAS.503.3065S}, spatial distribution \citep{2010A&A...521A..28A, 2012A&A...540A.123E, 2013MNRAS.436..275W, 2018MNRAS.478.5473L} or both redshift and spatial distributions \citep{2005ApJ...628L..97D, 2023A&A...669A.147B}. These properties were combined to produce dynamical state indicators for their clusters.

Other observational studies used a variety of other dynamical state indicators, such as the magnitude gap between the BCG and the second brightest galaxy \citep{2019ApJ...887..264R, 2020ApJ...904...36Z}, the center of mass offset between X-ray and luminosity weighted center \citep{2020MNRAS.497.5485Y, 2021MNRAS.504.5383D}, the concentration parameter defined by Navarro–Frenk–White profile \citep[][NFW]{1996ApJ...462..563N} fitting \citep{2019PASJ...71...79O}, and the detection of substructures using both positional and redshift information \citep{2023A&A...669A.147B}. 

Simulations were also used to investigate observational indicators in order to study the dynamical state of dark matter particles in clusters \citep{2012MNRAS.427.1322L, 2018MNRAS.478.4974L, 2021MNRAS.504.5383D, 2022MNRAS.516...26Z}. Moreover, some previous studies investigated other dynamical indicators that are only measurable in simulations \citep{2017MNRAS.464.2502C,2022MNRAS.514.5890L}. These indicators include virial ratio, 3-dimensional velocity dispersion deviation, and the satellite mass fraction including their dark matter mass component. 

Despite numerous attempts to employ various dynamical indicators, a consensus on measuring a cluster's dynamical state remains elusive. For example, some indicators disagree with other indicators, or have not had their usability verified in a quantified manner, and some simply cannot be directly applied to observations.

Recent studies have tried to solve some of these limitations by combining more than three indicators \citep{2020MNRAS.492.6074H, 2020ApJ...904...36Z, 2022MNRAS.514.5890L} or by setting up an intermediate zone for a transitional dynamical state \citep{2021MNRAS.504.5383D} instead of considering only mergers versus relaxed clusters. \citet{2022MNRAS.516...26Z} tried to provide the threshold-free function $\Lambda _{DS}$ using a three dynamical indicator combination from \citet{2017MNRAS.464.2502C}. However, in general, recipes were made using criteria with arbitrarily introduced boundaries for the definition of relaxed or unrelaxed, such as the Magnitude gap ($\Delta m_{12,r} \geq 2$ : the r-magnitude difference between BCG and the second brightest galaxy). Furthermore, the transitional parameter range was considered by excluding relaxed and unrelaxed ranges from the entire sample without a clear physical interpretation, and it remains difficult to apply these recipes directly to observational data.

To overcome these limitations, we consider impact on multiple dynamical state indicators over a complete range of mergers stages, from first infall, through final coalescence and eventually to several gigayears after coalescence. Mergers are energetic events that perturb the cluster dynamical state \citep{2006A&A...456...23B, 2019MNRAS.487.3922L, 2022MNRAS.516...26Z, 2022MNRAS.514.5890L}. There are two main factors that impact on the rate at which objects pass through the merger stages. The merger mass ratio is a key regulating factor that affects the time scale for coalescence and the strength of the collision \citep{2012A&A...545A..74G}. The relaxation time scale (or time since merger) is also an important factor in deciding the dynamical state of the galaxy cluster \citep{2022MNRAS.516...26Z, 2022AJ....164...95S}. However, these factors are difficult to constrain observationally due to projection effects.

%introduce our new approach
Therefore, this study aims to develop new observationally-applicable methods for measuring the dynamical state of galaxy clusters, derived based on simulation data. In the simulations, we can define the exact start of the merger event, measure the merger mass ratio, and trace the evolution through the entire merger process. To create the recipes, we will use dynamical indicators related to the optical properties of the cluster galaxies. These indicators allow us to build up a more extensive statistical sample than indicators using other wavelengths. This new method will not provide discrete division criteria but, instead, provide formulas with continuous reliability values. In this way, they are flexible according to a user's needs. We will also apply our recipes to observational data in order to demonstrate their ease of usage.

%layout of paper
The layout of this paper is as follows: we introduce our simulation data and sampling methods in Section \ref{sec:data}. We explain the observational indicators, the new method to combine them, and the two approaches to quantify the success of the recipe in Section \ref{sec:method}. In Section \ref{sec:result}, we show the responses of individual indicators to mergers, present the new recipes, and demonstrate their application to observational results. In Section \ref{sec:discuss}, we discuss comparisons between our and previous studies' results and present some notes on using our method. We conclude our study in Section \ref{sec:conclusion}.
The following cosmological parameters are assumed throughout this paper: $\Omega_{m} $= 0.3, $\Omega_{\Lambda}$= 0.7, $\Omega_{b}$= 0.047, and h = 0.684.

\section{Data} \label{sec:data}
\subsection{N cluster run simulation} \label{sec:sim}
The \texttt{N$\_$cluster run} simulation is a cosmological N-body dark matter$-$only simulation of a 120 Mpc $h^{-1}$ length cubic box. It has 169 snapshots with 100Myr time resolution from z=200. 
The particle mass is $1.07189\times10^9 M_{\sun}h^{-1}$. 
This simulation was conducted using the \texttt{Gadget3} \citep{2005MNRAS.364.1105S} code for cosmological N-body/SPH simulations. The six-dimensional friends-of-friends algorithm, \texttt{ROCKSTAR} halo finder \citep{2013ApJ...762..109B}, is used to define the halos of galaxies and clusters. Because it uses both 3D positional information and 3D velocity information, it can reliably distinguish satellite halos at the center of large halos and follow the merger process. The minimum virial mass of a halo that can be detected is $\sim 2 \times 10^{10}M_{\sun}h^{-1}$ $(N_{DM} = 20)$.

The stellar mass–halo mass relation ($M_{star}-M_{halo}$) is decided with Abundance matching \citep{2013ApJ...770...57B}. For satellites, tidal stripping of stars is considered using the $M_{star}-M_{halo}$ relations from \citet{2016ApJ...833..109S}. In this way, we can paint a stellar component onto the dark matter halos in our simulation, and the method can even account for the growth of the stellar component as the dark matter halo grows, and the tidal mass loss of stars in dense environments. These processes are required to successfully reproduce the observed masses of galaxies in a DM-only simulation. 

Many hydrodynamic cosmological simulations tend to form BCGs that are excessively massive \citep{2018MNRAS.475..648P, 2020MNRAS.498.2114H, 2020MNRAS.495..686A}. The abundance matching approach used here ensures reasonable BCG masses, which is important for calculating some of the dynamical indicators used in this work. The simulation dataset that we analyse was also used in the following works: \citep{2021ApJ...912..149S, 2022AJ....164...95S, 2022ApJ...934...86S, 2022ApJ...925..103C, 2022ApJ...940....2J, 2022ApJ...935...71K, 2022ApJS..261...28Y, 2023ApJ...943..148C}.
Further details of the simulation can be found in these papers and on the publically available data archive page \footnote{https://data.kasi.re.kr/vo/N$\_$cluster$\_$run}.

\begin{figure}[ht]
\centering
\begin{center}
\includegraphics[scale=0.75]{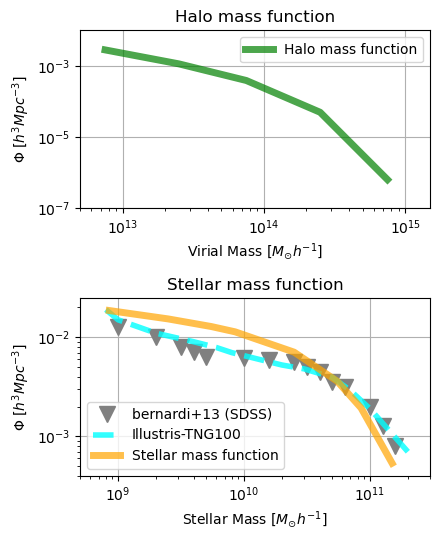}
\caption{(Upper) The halo mass (virial mass) function of the simulations used in this study. (Lower) The stellar mass function of subhalos using stellar masses derived from the abundance matching on the N$\_$cluster simulations (orange solid curve) and comparing with the stellar mass functions of the Illustris-TNG 100 simulation \citep{2018MNRAS.475..648P} (cyan dashed), and the SDSS data points from \citet{2013MNRAS.436..697B} (gray upside-down triangles).}\label{fig:massfunc}
\end{center}
\end{figure}

From this simulation data, we use 418 cluster halos ($M_{vir}>10^{14}M_{\odot}h^{-1}$) in 10 individual simulation volumes (halo mass ($M_{halo}$) is equivalent to $M_{vir}$ throughout this paper.). Three lines-of-sight (along the x, y, and z axes) are considered in order to include line-of-sight effects that can impact on the observed dynamical state indicators. Thus, the total number of cluster halos is 1254 (418$\times$3 lines of sight). 
Figure \ref{fig:massfunc} shows the halo mass function and the subhalo (galaxy) stellar mass function for the N$\_$cluster run data. The stellar mass functions include the SDSS (observational) and the Illustris-TNG 100 (simulation) results for comparison. While the general shape of the stellar mass functions is broadly quite similar, we note that the small upturn seen in the SDSS data for masses below $\sim 10^{10} M_{\sun}h^{-1}$ is not visible in the simulations. This is likely in part a result of the low number statistics in our sample compared to the statistics available in the SDSS sample, but we do not expect it to have a substantial impact on our main conclusions.

\subsection{Merger types and stages} \label{sec:samp}
We initially categorize the clusters using two quantities; the merger mass ratio and the frequency of mergers occurring prior to the merger. After that, the merger stage is defined using the time since infall of the merging halo.

First, all halos having masses $\geq 10^{14}M_{\sun} h^{-1}$ and their satellites were traced in a short redshift range (from redshift 0 to 0.5, $\sim$5 Gyr period). This is done to reduce any redshift dependencies. The merger starting point is defined as when the center of the infalling merging halo passes the virial radius of the main halo as measured in 3D. 

If there is no merger larger than 1:10 mass ratio during the last 5 Gyr, this halo is classified as being in a `relaxed' state. If a merger with 1:5 mass ratio happened only once during the last 5 Gyr, the cluster is classified as a `single major merger'. If major mergers happen more than once during that time, the cluster is considered a `multiple major merger'. If there are mergers with a mass ratio higher than 1:10 and a mass ratio less than 1:5, the cluster is considered a `minor merger'. In order to build our recipes on the clearest (and most clean) cases, we only use the relaxed state as the control sample and the single major merger as the merger sample in this study. 

The merger stages of single major merger clusters were divided into three categories; `recent' (1 Gyr after a merger), `ancient' (more than 3 Gyr after a merger), and `entire' (the combination of the two previous categories). The `entire' category probably best represents the observational case, where a mix of recent and ancient mergers can be found. The `recent' merger stage best represents ongoing mergers, and the `ancient' category best represents past mergers. In some cases of past mergers, full coalescence of the two cluster dark matter halos may have occurred.

In general, we compare the control sample to the merger sample at three different merger stages (entire, recent, and ancient). Three line-of-sights (along the x, y, and z axes of the simulation) are used.

\section{Method} \label{sec:method}
\subsection{Indicators} \label{sec:indproj}

To develop recipes for classifying the cluster dynamical states, we used five dynamical indicators. These are calculated on the projected plane for the three line-of-sights (along the x, y, and z axes) in order to include observational projection effects. Indicators were chosen that are observable, that can also be extracted from the simulation, and consider the cluster properties from their centers to their outskirts. 

\textbf{- sparsity : }
\begin{equation}
S= \Sigma M_*(r_{100})/\Sigma M_*(r_{50})
\end{equation}
where $\Sigma M_*(r_{100})$ and $\Sigma M_*(r_{50})$ are total stellar mass within $100\%$ and $50\%$ of the virial radius of cluster, respectively. This radius was chosen to avoid extreme values of sparsity to arise during a merger. This indicator is affected by how concentrated the bright galaxies are in the center of the galaxy cluster. The centrally concentrated systems will have a lower value. Thus, a lower value is expected for a relaxed cluster. This indicator uses projected distances to sum the stellar mass of a galaxy, and so is sensitive to mergers on the plane of the sky.

\textbf{- stellar mass gap : }
\begin{equation}
\Delta M_{*, 12}=M_{*, 2nd bcg}/M_{*, bcg}
\end{equation}
where $M_{*, bcg}$ and $M_{*, 2nd bcg}$ are the stellar mass of the BCG and second brightest cluster galaxy, respectively. This indicator essentially highlights the dominance of the BCG in the galaxy cluster. A BCG dominant system implies a more relaxed cluster dynamical state (i.e., a lower $\Delta M_{*, 12}$ value). Usually, the observational definition of magnitude gap finds the second brightest galaxy within half the virial radius of the cluster \citep{2003MNRAS.343..627J}. However, we used the entire cluster radius range because, otherwise, the indicator did not respond to the very recent mergers. This indicator is only very weakly affected by projection effects.

\textbf{- center offset : }
\begin{equation}
d_{off}=|P_{BCG}-P_{M_*\_weighted}|/r_{vir}
\end{equation}
where $P_{BCG}$ is the position of the BCG, and $P_{M_*\_weighted}$ is the stellar mass-weighted center of the cluster member galaxies. If the distance offset between the two centers is small, the system is relaxed. As the offset is measured on the plane of the sky, this indicator is sensitive to mergers on the plane of the sky. This indicator is normalized by the cluster's virial radius to allow for a comparison between clusters of different sizes.
\newline
\textbf{- velocity dispersion deviation : }
\begin{equation}
\xi=\sigma/\sigma_{t}, 
\end{equation}
where $\sigma$ is the projected velocity dispersion of the galaxy cluster, and $\sigma_{t}$ is the theoretical orbital velocity (calculated down our line-of-sight).
\begin{equation}
\sigma_{t}=\sqrt{GM_{total}/R}.
\end{equation}
This indicator shows how relaxed and symmetrical the line-of-sight velocities of member galaxies are. An approximate value of $\xi$=1 is expected for a relaxed cluster. We followed the formula of \citet{2017MNRAS.464.2502C}. As this indicator uses line-of-sight velocities, it is sensitive to mergers occurring down our line-of-sight (most others respond to mergers on the plane of the sky).

\textbf{- satellite stellar mass fraction : } 
\begin{equation}
f_{M_*}=M_{*, sat}/M_{*, cl}
\end{equation}
where $M_{*, sat}$ and $M_{*, cl}$ indicate satellite stellar mass and cluster stellar mass, respectively. This indicator shows how much of the total stellar mass of the galaxy cluster is in satellites. Relaxed clusters are expected to have smaller satellite stellar mass fractions. We followed the same formula given in \citet{2017MNRAS.464.2502C}, but they mentioned a caveat that relaxed clusters in observations tend towards values close to (but not exactly) one. This indicator is not expected to be highly affected by the direction of the merger.

A brief summary of all indicators is given in Table \ref{tab:ind}. Additional notes on how to use the indicators are provided in Section \ref{sec:notes}.

\begin{table}[ht!]
\centering
 \caption{Information of 5 dynamical indicators used.}\label{tab:ind}
 \renewcommand{\tabcolsep}{1.5mm}
  \begin{tabular}{cccc} 
  \hline \hline
    Name & symbol & plane & Obs type \\
 \hline  \hline 
sparsity & S & plane of sky & Photo\\
\hline
stellar mass gap &  $\Delta M_{*, 12}$  & plane of sky & Photo\\ 
\hline
center offset  &  $d_{off}$ &  plane of sky & Photo\\ 
\hline
velocity dispersion &  $\xi$ &  line of sight & Spectro\\ 
deviation  & & & \\  
\hline
satellite stellar&  $f_{M_*}$ & plane of sky & Photo \\
 mass fraction & & & \\
   \hline \hline
   \end{tabular}
 \tablecomments{ From left to right, columns show the name of the indicator, symbol, effective plane, and the observational data type (photometric or spectroscopic). }
\end{table}

Indicators are calculated for every 100 Myr time steps of the simulation. After the merger began, the different merger stages were compared in 1 Gyr bins since the start of the merger. The response of each indicator to a merger is different, and they have different dependencies on the line of sight. Thus, combining the indicators together is expected to provide additional information on the merger stage, beyond considering them individually. A detailed interpretation is given in Section \ref{sec:1d}.

\subsection{Rotation matrix} \label{sec:rm}
\begin{figure}[htpb]
\begin{center}
\includegraphics[scale=0.48]{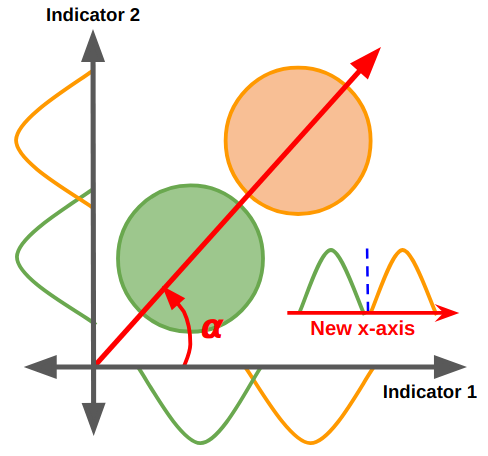}
\caption{Schematic explanation of the rotation matrix approach. The green and orange circles indicate mergers and control samples. The two Gaussian shapes on the \textit{x} and \textit{y}-axis represent one-dimensional indicator distributions of each sample. Suppose we rotate the \textit{x} as $\alpha$ angle, the new \textit{x}-axis is illustrated by the red arrows.}\label{fig:schematic}
\end{center}
\end{figure}

The rotation matrix is a transformation matrix for rotating coordinates. We start with indicator distributions in an indicator space where each axis is an individual indicator. First, we must standardize the indicators because they all have different units. 
\begin{equation}
\label{eqn:norm1}
    A=(I_1-I_1')/N_1 ,
\end{equation}
\begin{equation}
\label{eqn:norm2}
    B=(I_2-I_2')/N_2
\end{equation}
where $I_1$ and $I_2$ are indicators, $I'$ and $N$ are the mean and dispersion of $I_1$ and $I_2$. 

Next, we search for rotation angles that allow us to most cleanly divide the merger and control samples along the new axis after rotation (see Figure \ref{fig:schematic}). For example, in the case of a 2-dimensional rotation matrix, the formula is as follows.
\begin{equation}
\label{eqn:rm}
R=
\left[
\begin{array}{cc}
   cos\alpha & -sin\alpha \\
   sin\alpha &  cos\alpha \\
\end{array}
\right] 
\end{equation}
where $\alpha$ is the rotation angle rotating counterclockwise from point (0,0).

Then, we do rotation matrix multiplication with the standardized indicators, A and B. 
\begin{equation}
\left[
\begin{array}{c}
   X_{rot} \\
   Y_{rot} \\
\end{array}
\right] 
=R
\left[
\begin{array}{c}
   A \\
   B \\
\end{array}
\right] 
\end{equation}
We use the x-directional axis as an example. If we make a full formula with equations \ref{eqn:norm1} and \ref{eqn:norm2}, the final form of the new $x$-axis is given below.
\begin{equation}
\label{eqn:full}
   X_{rot} = cos \alpha A - sin\alpha B
\end{equation}
As shown in Figure \ref{fig:schematic}, if we rotate the axes by $\alpha$ degree from 0 degrees, the green and orange distributions are now fully separated along the new \textit{x}-axis, which is a linear combination of the two indicators (=equation \ref{eqn:full}). 

We can conduct a similar analysis in higher dimensional space, involving more than two indicators. We test every combination of indicators for 2, 3, and 4 indicators in order to search for the combination of indicators that best separates the merger and control sample along the new x-axis after rotation. We tried varying the bin size of the rotation angle from 1 to 30 degrees. We finally chose 10 degrees as the standard bin size in order to reduce computational time and to reduce local maxima problems. We test the success of each rotation angle at separating the control and merger samples using two quantifying methods described in the following section.

\subsection{Quantifying success : the success rate and the overlap percentage} \label{sec:quanti}

\begin{figure}[htpb]
\begin{center}
\includegraphics[scale=0.27]{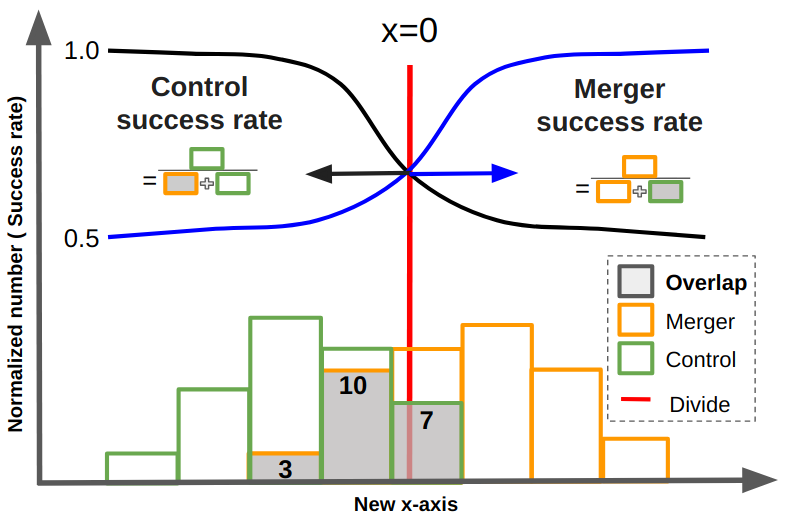}
\caption{Schematic explanation of the two methods used to quantify the success of a recipe at separating the control and merger samples. The green and orange colors represent the control and merger samples, respectively. The shaded histogram shows an overlapped range of the histogram. The amount of overlap is the one way to quantify the success. The second way is to measure what fraction of the galaxies are mergers (or control) at different locations along the new x-axis. The red line shows the chosen dividing point on the axis. Black and blue lines represent the success rate at finding clusters that are members of the control or merger samples.}\label{fig:quanti}
\end{center}
\end{figure}

To quantify the best combination result, we used `the success rate' and `the overlap percentage' (see Figure \ref{fig:quanti}).

\textbf{-Success rate:} The success rate quantifies the success of separating merger samples from control samples along the new \textit{x}-axis. The success rate of merger ($r_{succ,merg}$) and control ($r_{succ,cont}$) are represented by the following formula:
\begin{equation}
r_{succ,merg} = \frac{N_{M}(x_{rot}>x_0)}{N_{C}(x_{rot}>x_0)+N_{M}(x_{rot}>x_0)},
\end{equation}
\begin{equation}
r_{succ,cont} = \frac{N_{C}(x_{rot}<x_0)}{N_{C}(x_{rot}<x_0)+N_{M}(x_{rot}<x_0)},
\end{equation}
where $N_{M}$ and $N_{C}$ mean the number of clusters in the merger and control sample respectively. $x_{rot}$ is the x value in the new x-axis and $x_0$ is the dividing point along that new x-axis. 

This formula gives us continuous success rate values along the axis. If we measure the success rate at a certain position on the $x$-axis (e.g., $x$=0 in Figure \ref{fig:quanti}), the measured success rates are the fraction of merger (control) in the sample larger (smaller) than the position on the $x$-axis. A high success rate then means we can expect to detect a large and well-understood fraction of mergers at that location along the new x-axis.

For example, if at $x$=0, the merger success rate is 0.75, then 75\% of the clusters located at $x>0$ are mergers. In this way, the user can measure how likely their clusters are to be the mergers as a function of their location along the $x$-axis. Thus, we can interpret this value as the reliability of an individual cluster's classification. 

It is not necessary to choose a single location along the x-axis to use our recipes. However, for simplicity, we prefer to have just a single value for measuring the success of an individual recipe in order to compare different recipes. In this case, we choose $x\_0 = 0$  because recipes were standardized to be centered as 0. Nevertheless, we note that the continuous value of success rate along the new x-axis maybe great value to observational studies, without the need to limit to a single criterion. 

\textbf{-Overlap percentage:} The overlap percentage attempts to quantify the separation of the two sample histograms by measuring the fractional amount of overlap of their two normalized histograms. This is measured by the overlap percentage of the total objects in the sample (e.g., see grey histograms in Figure \ref{fig:quanti}). In this case, a small amount of overlap means the recipe is doing a better job at separating the control and merger samples. 

\begin{equation}
p_{overlap}=\Sigma N_{lower, i}/N_{total}
\end{equation}
where $N_{lower, i}$ is the the number of overlapping clusters between the control and merger sample histograms (see Figure \ref{fig:quanti}, the gray-shaded area shows $N_{lower, i}$ in each bin). In practice, these bars are chosen by selecting the smaller bar between the green and orange histogram bin in the region where they overlap. Each histogram is normalized by the smaller number of samples to consider the number of samples equally. $N_{total}$ is the total count of one sample. The number of bins for the histogram is calculated with $\sqrt {N_{total}}$ in order to take into account the sample size. Unlike with the success rate measurement, each recipe provides only a single value of the overlap percentage. 

For example, let's make the assumption that Figure \ref{fig:quanti} contains 64 merging clusters and 64 control clusters. Then, the total number of samples in a single histogram, $N_{total}$ is 64, and the number of bins will be 8. There are only three bins overlapped in the middle of two histograms, with (from left to right) $N_{lower, i}$= 3, 10, and 7 clusters shared between the two histograms. Then, $\Sigma N_{lower, i}$ is 20, and $p_{overlap}$ will be 0.31.

Using these two different methods to quantify the success of a recipe, we selected the best recipes as having a combination of a higher success rate and lower overlap percentage. The use of two different quantifying methods helps us to avoid false measurements caused by outliers. The two quantitative values were calculated at for each of the 10-degree angles of rotation applied in multi-indicator dimension space. As quantified by these two methods, the success of the best recipes that we find is provided in Section \ref{sec:result}.  

\section{result} \label{sec:result}

In this section, we will begin by showing individual indicators' responses to mergers. We then describe our new recipes, which are made by combining them together. Finally, we will demonstrate the ease of use of our new recipes on an observational data set.

\subsection{ 1D parameter tendency } \label{sec:1d}
\begin{figure*}[htpb]
\begin{center}
\includegraphics[scale=0.53]{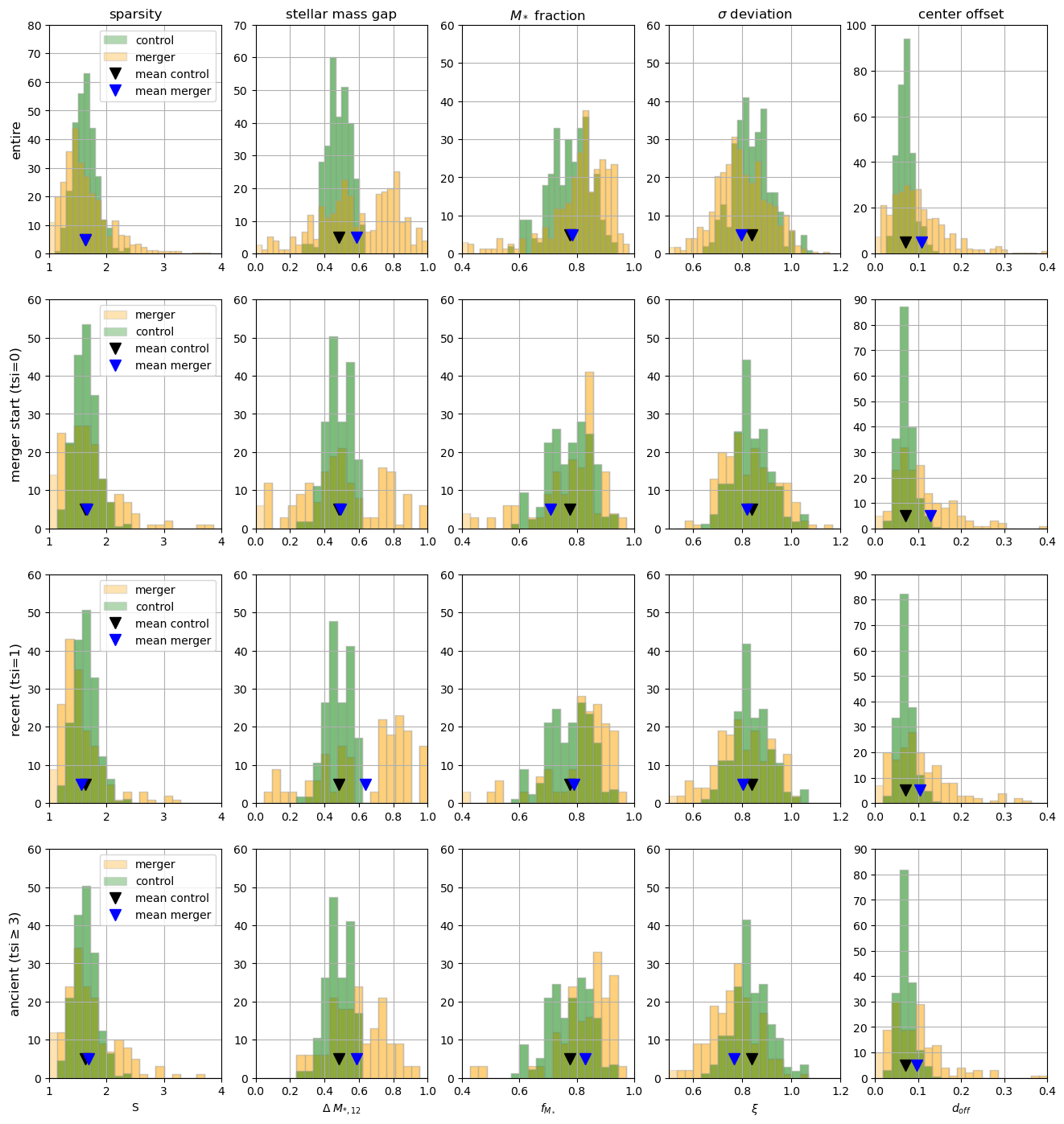}
\caption{Histograms of change of each dynamical indicator. In the horizontal direction, panels show the sparsity, stellar mass gap, satellite stellar mass fraction, velocity dispersion deviation, and center offset. In the vertical direction, panels show the entire merger sample, merger start (tsi = 0 Gyr), recent (tsi = 1 Gyr), and ancient (tsi $\geq$ 3 Gyr) merger sample (tsi means the time since infall). The green and orange histograms represent the relaxed (control) and unrelaxed (merger) dynamical state samples. Black and blue up-side-down triangles represent the mean values of each histogram.  }\label{fig:histo}
\end{center}
\end{figure*}

Figure 4 shows the histogram distribution of each individual indicator over time since merger (tsi). The figure consists of multiple panels where each column represents the distribution of a specific indicator as it evolves over time since the merger. Green and orange histograms represent the relaxed and merger sample distributions, respectively. The upside-down triangle symbols represent the mean value of each distribution. In this figure, we see the response patterns and effectiveness of individual indicators in detecting mergers at different stages of the merger process.

Most indicators have quite overlapped histograms between the control (green) and merger samples (orange), which demonstrates that the use of a single indicator may not be highly effective for detecting mergers. For the 'entire' merger sample, the velocity dispersion deviation indicator returns 70.26$\%$, the highest overlap percentage (least successful at separating mergers and control), and the stellar mass gap indicator returns 40.87$\%$, the lowest overlap percentage (most successful).

We also note that each indicator responds to mergers in a different way, and the response can occur at slightly different times.

The sparsity (S) indicator can show both increases and decreases in its value in response to different merger stages. While the secondary halo falls into the main cluster, a large amount of stellar mass associated with the secondary halo could be found away from the cluster's center, enhancing the sparsity. Later, when the secondary halo reaches the cluster's mass center (e.g., near tsi=1 Gyr), the concentration of mass near the cluster center may cause a reduction in the sparsity of the merger sample. After the secondary passes the center, larger values of sparsity reappear as the secondary separates from the main cluster after the first passage. Consequently, this indicator can exhibit fluctuating values throughout the merger stages. 

Meanwhile, the stellar mass gap ($\Delta M_{*, 12}$) reacts to the merger quickly (tsi=1 Gyr) as soon as the secondary enters the cluster. It then remains perturbed for a long time, only returning to more relaxed values after the secondary main galaxy has been tidally disrupted or merged with the primary cluster. 

The velocity dispersion deviation ($\xi$) steadily increases with increasing time since the merger. This could make it a useful indicator for detecting mergers that happened long ago. This indicator can be interpreted as a detection of the disturbed velocity signatures of substructures in the cluster. When an infalling secondary halo approaches the main cluster, the velocity dispersion deviation will differ from the value of the relaxed sample (in our simulation, this value is approximately 0.8, see Section \ref{sec:notes}).

The center offset ($d_{off}$) and satellite stellar mass fraction ($f_{M_*}$) react quickly to the merger (tsi=0 Gyr) and remain perturbed for a long time, again making the potentially useful for detecting ancient mergers. This is due to the presence of substructures, which require several cluster relaxation times to be disrupted.

While the stellar mass gap ($\Delta M_{*, 12}$) proves to be a valuable individual indicator for detecting mergers, we expect that combining multiple indicators will lead to even greater effectiveness. This is because various indicators exhibit distinct responses at various stages of the merger process and are influenced differently by projection effects. Therefore, by combining these indicators, our aim is to exploit their diverse strengths and sensitivities to different merger timescales. Furthermore, to effectively address the projection effects, we anticipate to observe cluster mergers occurring across different orientations, including those observed directly on the sky, along our line of sight, and at various angles in between. Thus, we test the effectiveness of recipes consisting of different combinations of indicators in various configurations, testing their ability to detect both recent and ancient mergers.

\subsection{ New recipes for determining cluster dynamical state by combining multiple indicators } \label{sec:best}
We combined 2, 3, and 4 indicators together in various combinations, with the goal of searching for the most effective recipes to determine the dynamical state of galaxy clusters observationally (Table \ref{tab:bestrecipes}).
Table \ref{tab:bestrecipes} lists the 9 best recipes and their indicator combination, coefficient, and two quantifying values with dividing points. Each combination is separated by the merger stage (entire, recent, and ancient) and the number of indicators involved in the recipe (2, 3, and 4). The Table lists the mean values of the three lines of sight (along the x, y, and z axes), and the uncertainties present the maximum and minimum values. Figure \ref{fig:2exp} shows the results of applying a combination of two indicators. Figures in Appendix \ref{sec:allrecip} present the results of all our configurations.

\textbf{- Dominant indicator :} Table \ref{tab:bestrecipes} shows that the most common indicators among the best 9 recipes are the stellar mass gap ($\Delta M_{*,12}$) and the center offset ($d_{off}$). Indeed, all recipes have the stellar mass gap as a base component, and this indicator has the largest coefficient values in most recipes, meaning it is dominant. The center offset is the next most common indicator to appear. For detecting ancient mergers, the satellite stellar mass fraction appears to be the next most influential indicator after the stellar mass gap. It is because of how long that indicator remains perturbed following a merger, as could also be seen in the histograms of the individual indicators.

We anticipated a consistent pattern of indicator selection as we introduced more combined indicators, including the addition of new indicators to the existing combinations. However, in practice, selecting the optimal recipes seems somewhat random because there are sometimes only small differences in success rate when testing the different combinations. This is especially the case for recipes where the center offset, satellite stellar mass fraction, and velocity dispersion deviation are interchanged.

%%======================= table ===============================

\begin{table*}[t]
\caption{Best combination results using 2, 3, and 4 indicators}\label{tab:bestrecipes}
\centering
\begin{tabular}{ccccccccc} 
\hline
Number of&Merger & Indicator &  Coefficient & Dividing &  Success & Overlap \\
combination &stage & (A,B,C,D) &   & point &  rate & percentage \\

\hline   
&Entire& $\Delta M_{*, 12}$, $d_{off}$ &  -0.87$ A$+0.5$ B$& $0.09^{+ 0.08}_{-0.08}$ & $0.81^{+ 0.02}_{-0.03}$ & $40.34^{+ 1.98}_{-1.19}$ \\
2 indicator&Recent& $\Delta M_{*, 12}$, $d_{off}$ &  -0.98$ A$-0.17$ B$& $-0.09^{+ 0.19}_{ -0.19}$ & $0.83^{+ 0.02}_{-0.04}$ & $30.97^{+ 2.70}_{-1.97}$  \\
&Ancient& $\Delta M_{*, 12}$, $f_{M_*}$ &  0.68$ A$-0.34$ B$& $0.26^{+0.16}_{-0.00}$ & $0.78^{+ 0.02}_{-0.04}$ & $45.08^{+ 0.90}_{-0.84}$  \\

\hline 
&Entire& $\Delta M_{*, 12}$, $\xi$, $d_{off}$ &  -0.98$ A$+0.17$ B$+0.03$ C$& $0.02^{+ 0.14}_{- 0.14}$ & $0.81^{+-0.01}_{-0.01}$ & $38.92^{+ 1.44}_{-2.45}$  \\
3 indicator&Recent& $\Delta M_{*, 12}$, $f_{M_*}$, $d_{off}$ &  -0.98$ A$+0.16$ B$+0.06$ C$& $0.07^{+ 0.13}_{- 0.13}$ & $0.85^{+ 0.03}_{-0.03}$ & $27.95^{+ 1.75}_{-0.94}$ \\
&Ancient& $\Delta M_{*, 12}$, $f_{M_*}$, $\xi$, &  0.77$ A$+0.63$ B$-0.11$ C$& $0.16^{+ 0.37}_{- 0.02}$ & $0.78^{+ 0.00}_{-0.01}$ & $38.67^{+ 4.54}_{-2.29}$  \\

\hline 
&Entire& $\Delta M_{*, 12}$, $f_{M_*}$, S, $d_{off}$ &  0.80$ A$+0.29$ B$-0.15$ C$+0.5$ D$& $-0.11^{+ 0.04}_{-0.22}$ & $0.82^{+ 0.01}_{-0.01}$ & $39.17^{+ 0.82}_{-1.61}$ \\
4 indicator&Recent& $\Delta M_{*, 12}$,  $\xi$, S, $d_{off}$ &  -0.62$ A$+0.11$ B$+0.11$ C$-0.77$ D$& $-0.08^{+ 0.18}_{- 0.31}$ & $0.84^{+0.02}_{-0.02}$ & $25.94^{+ 2.22}_{-1.97}$  \\
&Ancient&  S, $\Delta M_{*, 12}$, $f_{M_*}$, $\xi$ & 0.09$ A$-0.55$ B$-0.66$ C$+0.50$ D$& $0.00^{+ 0.17}_{- 0.20}$ & $0.78^{+ 0.03}_{-0.02}$ & $35.53^{+ 3.42}_{-2.01}$  \\
\hline 
\end{tabular}
 \tablecomments{Table shows the best recipes divided by merger stage and the number of indicators combined in each recipe. A to D in the coefficient column are standardized \textit{n}-th indicator listed in the indicator column. The dividing point is the best separation point on the histogram of each sample on the new $x$-axis (axis1) of Figure \ref{fig:2exp}. The value of the success rate provided is the mean of the success rate for the merger sample and the control sample. Uncertainties show the maximum and minimum values from the three lines of sight (along the x, y, and z axes).}
\end{table*}
%%================================= table =========================================

\begin{figure*}[ht!]
\centering
\begin{center}
\includegraphics[scale=0.65]{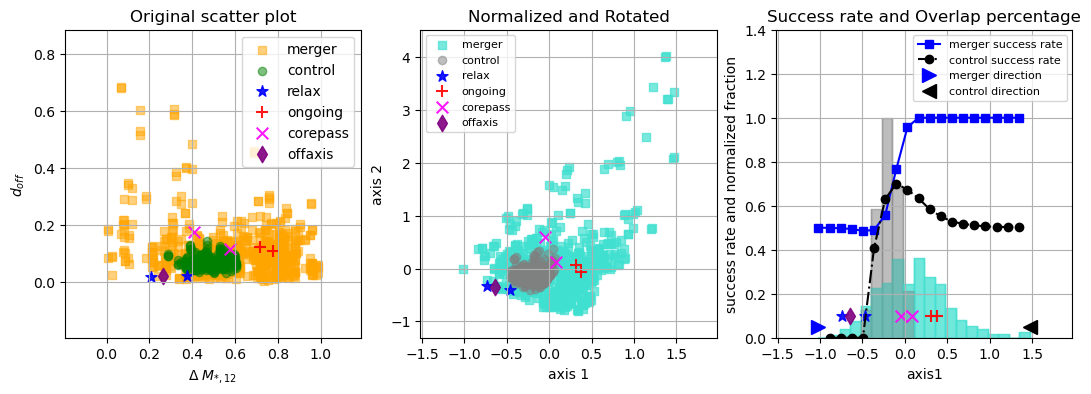}
\caption{Example of an application of the 2 indicator best recipe using the entire merger stage sample. The left panel is the 2-dimensional scatter distribution of the first and second indicators. The middle panel shows the standardized and rotated (converted) scatter distribution of indicators. Axis 1 and 2 on this figure are linear combinations of $\Delta M_{*, 12}$ and $d_{off}$ (see Table \ref{tab:bestrecipes}). The right panel shows histograms of each sample on the new $x$-axis (=axis 1) and line plots of the success rates. Left or right-sided triangles point to the measuring direction of the success rate for each sample. See legends for the symbol meanings. The relaxed and off-axis symbols are discussed in Section \ref{sec:appobs}.}\label{fig:2exp}
\end{center}
\end{figure*}

\begin{figure}[ht!]
\begin{center}
\includegraphics[scale=0.80]{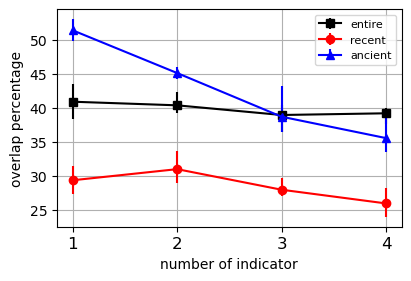}
\caption{The change in the overlap percentage of the control and merger sample distributions as a function of the increasing number of indicators combined together in the recipe. Black, red, and blue lines represent the entire, recent, and ancient merger samples respectively. Error bars show the maximum and minimum values from three lines of sight (along the x, y, and z axes).}\label{fig:comp}
\end{center}
\end{figure}

\textbf{- Dependency on number of indicators :} We compare the success of the different recipes with varying numbers of indicators using the overlap percentage (see Figure \ref{fig:comp}). The overlap percentage is the best choice for this task because it provides a single value to quantify the success of each recipe. As shown in Table \ref{tab:bestrecipes}, we split the rows by the number of combined indicators. Although there are small fluctuations, the overlap percentage is overall reduced by the inclusion of more indicators, in most cases. 

Interestingly, the ancient merger sample shows a significant improvement (15$\%$ decrease) of the overlap percentage as we move from a recipe that uses one indicator (50$\%$) to a recipe that combines four indicators together (35$\%$). This shows the value of combining indicators for detecting those mergers that are more difficult to detect. We note that all overlap percentages in the ancient sample are higher than those in the recent sample. However, the inclusion of additional indicators significantly narrows this gap between the two samples. The recent merger sample exhibits a smaller improvement (5$\%$ decrease) from 2 indicators to 4 indicators. There is a slight increase of overlap from 1 indicator to 2 indicators, but it is marginal within the error bars. The entire merger sample shows a steady decrease as the number of indicators increases, though only by a small percentage.

When considering the individual lines-of-sight, we noticed decreasing trends in the overlap percentages. However, when combining the three lines-of-sight, this trend is less prominent. Nonetheless, the clear indication of decreasing overlap in the ancient merger sample is an encouraging result for our approach. It indicates that combining multiple indicators can effectively separate merger and control samples even a long time after the merger.

\textbf{- Usage of recipe :} Figure \ref{fig:2exp} shows an example of applying a 2 indicator combination recipe for the entire merger stage sample. The scatter plot in the left panel shows the sample distribution in 2-dimensional parameter space. The merger sample has a very scattered distribution; the control sample is much more concentrated at low offset and low stellar mass gap values. (The other symbols will be discussed in Section \ref{sec:appobs}).  

In this case, we set each cluster's measured stellar mass gap as the $I_1$ value and the measured center offset as the $I_2$ value in the 2 indicator combination recipe for the entire merger sample (i.e., the first row of Table \ref{tab:bestrecipes}). From this, we obtain the rotated positions of cluster samples on the new $x$-axis in the middle panel of Figure \ref{fig:2exp}.

The middle panel represents standardized and rotated (converted) sample distributions. Sample colors were changed to illustrate that samples have already been rotated. The distribution is reorganized based on the 0 value along the $x$-axis. We emphasize that our recipes focus only on the position of clusters along the $x$-axis as a means to determine the dynamical state. Nevertheless, we show both axes on the plots so that the rotation that has been applied can be seen by eye, for ease of understanding.

Using histograms in the right panel of figure \ref{fig:2exp}, the right panel exhibits how well separated the two samples are on the $x$-axis. The success rate lines can be used to quantify the reliability of the dynamical state measurement of a sample of clusters as a function of their position along the $x$-axis. For example, if a galaxy cluster is located on the right side of zero (x$>$0), this cluster can be interpreted as being a merging cluster with 100$\%$ reliability, according to the fraction represented by the blue line.

\subsection{Demonstration of application to observation data} \label{sec:appobs}

We applied our new recipes to observational clusters to demonstrate the ease with which they can be applied. For the data, we choose the merging cluster collaboration \citep[$MC^2$:][]{2019ApJ...882...69G} and the Hectospec cluster survey \citep[HeCS:][]{2013ApJ...767...15R}. The $MC^2$ provides clusters in various merger stages which have been classified according to various diagnostics that differ from our own approach. The HeCS is a useful survey for our approach as it has many spectroscopically confirmed member galaxies. We chose clusters belonging to both catalogs in order to obtain samples with high spectroscopically confirmed membership and an independent assessment of the merger stage. 

Next, we classify the cluster merger stages using our new recipes and compare them to the merger stages suggested by the $MC^2$ diagnostics. For example, the `ongoing merger' stage is defined by having entangled two clumps of luminosity contours in the elongated X-ray image (and radio) and having a broad single peak in the velocity histogram of the galaxies. Meanwhile, a `core-passing merger' has two prominent density contours and two (or more) prominent peaks in the galaxy member velocity distributions. The `off-axis merger' is defined based on a cool-core feature (cool gas temperature) and the presence of a radio relic, which are interpreted as a merger that occurred but did not impact on the center of the cluster. Relaxed clusters are chosen from literature \citep{2013MNRAS.436..275W, 2022ApJ...928..170K}.

As a result, we could choose 4 clusters for `relaxed' (2) and `ongoing' (2) merger stages. However, we choose an additional 3 clusters in the transitional dynamical phase to provide additional data points, such as `core-passing' (2) and `off-axis merger' (1), although we caution that these additional clusters have low spectroscopic completeness. A summary of information about the clusters can be found in Table \ref{tab:obssample}. 

\begin{table}[ht!]
\centering
 \caption{Example clusters in HeCS and MC$^2$ catalog.}\label{tab:obssample}
 \renewcommand{\tabcolsep}{0.7mm}
  \begin{tabular}{cccccc} 
  \hline
    Merger & Name & Redshift & Mass & Comple & Number \\
    type & & & [$10^{14}h^{-1}M_{\odot}$] & -teness & \\
 \hline   
Relaxed & A963 & 0.2041&4.01& 0.93 & 341\\
Relaxed &  A2261  &0.2242 &2.62  &  0.71 & 358\\
 \hline   
Ongoing &  A2061 & 0.0783& 5.27& 0.92 & 287\\
Ongoing &  A2255 & 0.0801&7.51 & 0.86 & 386\\ 
 \hline   
Core-pass&  A2443 & 0.1102& 1.89&   0.25 &  87\\
Core-pass&  A2034 & 0.1132 & 5.03&  0.29 &  175\\  
 \hline   
Off-axis &  A0115 & 0.1916 & 10.6&  0.35 &  140\\
   \hline
   \end{tabular}
 \tablecomments{References of A963 and A2261 are \cite{2013MNRAS.436..275W} and \cite{2022ApJ...928..170K}. Other cluster are from \cite{2013ApJ...767...15R}, \cite{2019ApJ...882...69G} and references therein. }
\end{table}

To compare the observational data with the simulation, observational magnitudes are converted to stellar mass using the conversion formula of \citet{2010MNRAS.404.2087B}. Notes on using equation 14 are provided in section \ref{sec:notes} (details can be found in section 2.4 of \citeauthor {2010MNRAS.404.2087B}).

\begin{equation}
\scriptstyle log_{10}\frac{M_{* Bell}}{M_\sun} = 1.097(g-r)-0.406-0.4(M_r-4.67)-0.19z
\end{equation}

We also convert redshift to line-of-sight velocity measured with respect to the center of the cluster. Then, we measure the five indicators, and insert them into the recipes shown in Table \ref{tab:bestrecipes}, and over-plot the results with symbols on the original data distributions (Figure \ref{fig:2exp}). In Figure \ref{fig:2exp}, square and circle symbols (merger and control samples) are measured from the simulated clusters, and the other symbols (relaxed to off-axis) are measured from the observational clusters.

We obtained good agreement between our merger and control sample distributions and the independent dynamical state measurements from the observational data. Relaxed and ongoing merger clusters are always located where they were expected in our control and merger sample distribution. 

As mentioned above, the success rate can be used as a reliability value to judge an individual cluster’s dynamical state. Following the merger success rate line (blue line in the right panel of Figure \ref{fig:2exp}), clusters located at a position of greater than zero on the new $x$-axis are classified as mergers with a 100$\%$ success rate, according to the simulated clusters. Conversely, the control success rate line (black line) shows the percentage of relaxed clusters found to the left of the position on the $x$-axis, which peaks at about 70$\%$ at a value of -0.4 on the new $x$-axis. 

From these success rate lines, we can demonstrate a broad agreement between the dynamical state measurements from our method and the $MC^2$'s method for estimating cluster dynamical states. For example, the ongoing merger samples are found in the position along the $x$-axis that signifies 100 $\%$ merger state according to our recipe. Meanwhile, the core passing merger sample is located along the $x$-axis where 80$\%$ of the clusters are expected to be mergers, according to our method (Figure \ref{fig:2exp}).

The core-passing clusters mostly fall at locations between the relaxed and ongoing merger clusters. However, an off-axis merger cluster shows more similar characteristics with relaxed clusters than any of the merging clusters. Most of the recipes show a significant separation of relaxed and merger samples at 0 on the new $x$-axis. 

The position of the data points of core-passing clusters seems to change according to the merger stage. The entire merger stage shows that most symbols separate well from each other. However, core-passing clusters overlap with the ongoing merger cluster in the recent merger stage and with the relaxed clusters in the ancient merger stage. This can be interpreted as showing that the recent merger recipes can successfully detect recent merger features.

Furthermore, the recipes combining 4 indicators together always seem to work better than others (see Appendix \ref{sec:allrecip}). This recipe can divide relaxed and core-passing merger samples even in the ancient merger stage sample and produces a clear separation among the observational symbols in the entire merger stage sample.

To summarize, all the results from the recipes seem to agree well with both observational and simulation data. This shows the applicability of our method to observational data.

\section{discussion} \label{sec:discuss}
In this section, we compare our results with previous research and note some precautions on the use of our method.

\subsection{Comparison with previous studies} \label{sec:prev}
\textbf{Dominant indicators :} \citet{2022MNRAS.514.5890L} also ranked important indicators for each number of combinations using simulation, optical, and X-ray observational data. Their highest-ranked indicators were mostly related to the mass. The virial ratio was their highest-ranked (most important) indicator, but it can not be directly measured observationally. There are 2 indicators that are similar to indicators used in this study, our `stellar mass gap ($\Delta M_{*, 12}$)' and their `stellar mass fraction near BCG ($f_{ste}$)' (in terms of using the stellar mass ratio) and our `center offset ($d_{off}$) and their center of mass offset ($\Delta_{r}$)'. The $f_{ste}$ and $\Delta_{r}$ were ranked in the second and fifth highest scores of dominance, respectively, in \citet{2022MNRAS.514.5890L}'s paper. 

\citet{2019ApJ...887..264R} and \citet{2020ApJ...904...36Z} also pointed to the magnitude gap and center of mass offset as the most important combination of indicators. \citet{2020MNRAS.492.6074H} used the satellite mass fraction, the center of mass offset, and the virial ratio for their relaxation parameter ($\chi$). 

Summarizing the previous studies, the center of mass offset is universally considered an effective indicator for simulations and observations, in common with our study. The virial ratio is generally considered the best indicator in simulations, and the magnitude gap seems to be the best indicator for observations, perhaps due to its insensitivity to line-of-sight effects. Similarly, in this paper, $\Delta M_{*, 12}$ is the most dominant indicator for recipes, and the relative dominance of indicators in the recipes seems broadly consistent with previous studies, where they can be fairly compared.

\textbf{Number of indicators for combinations :} How many indicators should be combined to detect mergers best? \citet{2022MNRAS.514.5890L} tested this question with various types and numbers of dynamical indicators based on a Random Forest machine-learning approach. In their Figure 6, the highest scores converged around when using 3 indicators combined (where a high score means the dynamical state was well described). 

We can not directly compare their result with our figure \ref{fig:comp} because of the difference in methods and data. However, the general tendency of the score seems similar to that shown in our Figure \ref{fig:comp}; The overlap percentage decreases until we reach 4 indicators, and the amount of reduction of the overlap percentage decreases gradually as less and less dominant indicators are included in the recipe.

\textbf{Division criteria :} In the introduction, we mentioned some papers that provided an intermediate zone, or a continuous criteria for the dynamical state. For instance, \citet{2022MNRAS.516...26Z} set the best threshold criteria as the cross-section of a double Gaussian fit to the indicator value distribution, differing from previous work that used existing criteria values. \citet{2021MNRAS.504.5383D} made a `hybrid' range for a continuous expression of dynamical states. This hybrid range includes all the exceptions from the criteria for the relaxed and unrelaxed states and shares some of the parameter ranges of the relaxed and unrelaxed states. 

In this study, we also provide a continuous success rate as a criterion for the dynamical state of the clusters. Users can measure the probability that a cluster is a merger, based on the cluster's position along the new x-axis. For example, we provide 0 as dividing points in Table \ref{tab:bestrecipes}, but the core passing merger sample is located near 0 in Figure \ref{fig:2exp}.  If required, a user can define their own choice of what probability must be considered in the merger category by using a continuous success rate.

\subsection{Some notes using our recipes} \label{sec:notes}

\textbf{Redshift and mass dependency :} \citet{2023MNRAS.519.6111V} showed the existences of redshift and mass dependencies of their dynamical state indicators. They tested a range of halo mass from $10^{12}$ to $10^{15} M_{\odot}h^{-1}$ within a redshift range from 5 to 0. Indicators showed a decrease with increasing redshift, and most of the indicators do not have mass dependencies.

We also checked our indicators for a dependence on redshift and mass. To reduce such dependencies, we limit the mass considered ($M= 10^{14} \sim 10^{15}M_{\odot}h^{-1}$) and also the redshift ($z= 0.5 \sim 0$) range. The maximum difference of an indicator by mass is about 20$\%$, and redshift dependency is less than 10$\%$, which is a small difference when compared to \citet{2023MNRAS.519.6111V}. 

Furthermore, we also combined various indicators and made recipes by dividing the merger stage, which might also have helped to reduce dependencies. Therefore, not only do we find that the dependencies are small for indicators, but also that our overall method is not highly affected by any mass and redshift dependencies that exist. We also plan to take into account the dependency on redshift and mass in our future work, as a potential further improvement to the recipes presented in this study.

\textbf{Dependency of velocity dispersion deviation indicator on the depth of spectroscopic data :} Using the cosmological simulations, we find that the velocity dispersion deviation is affected by the mass limit of galaxies considered. This is because the velocity dispersion of massive galaxies is smaller than that of low-mass galaxies, even in the same cluster. This is likely due, in part, to mass segregation. Therefore, when the velocity dispersion deviation parameter is measured for a cluster, it is important to consider the observational depth of the spectroscopic galaxy catalog. As a result, ideally, the mass limit of the observations should be quite similar to that of the simulations that were used to create the recipe.

Therefore, for the results where we compare with HeCS data, we used a similar depth as in that survey observational data, and removed dwarf-size stellar mass galaxies ($M_{*}< \sim10^{8}M_{\odot}$) from cluster members. This situation can have a significant impact on the velocity dispersion deviation indicator. Indeed, it could potentially affect other indicators as well, although it is expected to have a less major impact on others unless the mass cut is very severe.

\textbf{Stellar mass conversion with magnitude :} In section \ref{sec:appobs}, we used the magnitude to stellar mass conversion equation from \citet{2010MNRAS.404.2087B}. This equation uses the SDSS \texttt{cModel} magnitude and considers the K-correction and the evolution correction. However, \citet{2020ApJ...897...15T} reported that the SDSS \texttt{cModel} magnitude has about 0.3 magnitude difference by mass and redshift.

To investigate this issue, we compared stellar masses generated from the spectral energy density (SED) fitting of member galaxies of sample clusters \citep{2015yCat..22190008C} and stellar mass calculated from \citet{2010MNRAS.404.2087B}'s stellar mass conversion. We find that the resulting stellar masses from \citet{2010MNRAS.404.2087B} did not show a significant offset with the stellar mass of \citet{2015yCat..22190008C}. A fit to the one-to-one relation gives the following equation: y=0.9726x+0.2576.

\section{Conclusion} \label{sec:conclusion}

In this study, we develop new recipes for classifying the dynamical state of observed galaxy clusters. 

We use the N$\_$cluster simulation to design the recipes because in this way, we can control the types of merger and trace out the full merger evolution in the simulation. We divide the sample into a merger and a relaxed sample (referred to as the control sample) using the merger mass ratio and merger history of the clusters. For clarity of results, we select only single major mergers for the merger sample. Then, this sample is further divided into the recent and the ancient mergers to test our recipes for detecting mergers at different stages (see Section \ref{sec:data}).

We define and measure five observable dynamical indicators, S, $\Delta M_{*, 12}$, $d_{off}$, $\xi$, and $f_{M_*}$ (see Section \ref{sec:indproj} and Table \ref{tab:ind}). These indicators are measured along a line-of-sight, like in the observations. If we plot these indicators in a multi-dimensional space (one indicator per axes of the space), merger clusters will tend to be located in a different region in the multi-dimensional indicator space than the control clusters. 
A rotation matrix is then used to rotate the multi-dimensional indicator space until the control and merger samples best separate along the new x-axis after rotation. The angles of the rotation can then be used to derive the new recipe for measuring the dynamical state of the clusters. 

The success of each recipe is quantified by two measures, the success rate and the overlap percentage (see Section \ref{sec:quanti} for details). Continuous measurements of the success rate are provided along the new x-axis. In this way, based on the location of the observational sample of clusters on the new x-axis, users can assess what fraction of the clusters are mergers using the success rate. 

Our results can be summarized as follows:
\begin{itemize}

\item We provide recipes using 2, 3, and 4 indicator combinations for each merger stage sample (entire, recent, and ancient). Across all the recipes, the stellar mass gap is generally the dominant indicator. The second most dominant indicators are $d_{off}$. The other 3 indicators are found to have a relatively similar importance for measuring the dynamical state of the clusters. 

\item Increasing the number of indicators combined in the recipe leads to an improvement in the ability to detect dynamical states. In particular, the ancient merger sample shows the most prominent improvement with increasing numbers of indicators, which is exciting as ancient mergers are expected to be the most difficult to detect.

\item Our multi-indicator recipes can have high effectiveness at separating mergers from relaxed clusters. For example, our recipe that combines four indicators simultaneously has a success rate that is as high as about 80$\%$. In other words, from a sample of clusters about 80$\%$ are expected to be mergers. Meanwhile, the overlap between the merger and control sample histogram is only about 30$\%$ meaning the total sample is quite well separated by the recipe. 

\item We demonstrate the ease of application of our method to 7 observational cluster samples, which are selected from the $MC^2$ \citep{2019ApJ...882...69G} survey and the HeCS \citep{2013ApJ...767...15R} catalog data. The dynamical states of these clusters have been determined previously using an independent approach from our own recipes. When applying our recipes to this observational cluster sample, we find overall good agreement with the dynamical states that were previously determined (see Figure \ref{fig:2exp} and appendix \ref{sec:allrecip}). 

\end{itemize}
\vspace{-\topsep}

For future work, we will design recipes that consist of photometric-only indicators. Such recipes will be useful as they can be applied to large samples of observed clusters, giving us excellent statistics, such as clusters in the Sloan Digital Sky Survey (SDSS), and the Southern Photometric Local Universe Survey (S-PLUS). We will also design recipes that are tailor-made for specific surveys, including spectroscopic surveys, such as the Hectospec Cluster Survey (HeCS). Through these, we aim to build large publicly available catalogs of the dynamical states of clusters for both the northern and southern hemispheres. Our hope is that such catalogs will a valuable scientific resource, and of great value in improving our understanding of structural growth in the universe and its impact on galaxy evolution.

\begin{acknowledgments}
We thank the anonymous referee for the useful comments. We thank H.S. Hwang for allowing us to use HeCS data and J.-W. Kim and K.-I. Seon for helpful advice. Hectospec observations used in this paper were obtained at the MMT Observatory, a joint facility of the Smithsonian Institution and the University of Arizona. This research was supported by the Korea Astronomy and Space Science Institute under the R$\&$D program (Project No.2023-1-830-00), supervised by the Ministry of Science and ICT. J-.H.Shinn was supported by the Korea Astronomy and Space Science Institute under the R$\&$D program (Project No.2023-1-868-03), supervised by the Ministry of Science and ICT. H.K. was supported by the UST Overseas Training Program 2022 (No.202206) and UST Young Scientist+ Research Program 2022 (No.2022-YS-46) funded by the University of Science and Technology. K.W.C. was supported by the National Research Foundation of Korea (NRF) grant funded by the Korea government (MSIT) (2021R1F1A1045622). J.H.Shin. was supported by the National Research Foundation of Korea(NRF) grant funded by the Korea government(MSIT) (2022M3K3A1093827). J.Y. was supported by a KIAS Individual Grant (QP089901) via the Quantum Universe Center at Korea Institute for Advanced Study. 
\end{acknowledgments}


\vspace{5mm}



\appendix
\counterwithin{figure}{section}

\section{Higher dimensional rotation matrix} \label{sec:high-d}
For the 3-dimensional rotation matrix, fundamental rotation matrices are shown below. The rotation vector for three dimensions combines three fundamental rotation matrices using matrix multiplication.
\begin{equation}
R=R_{z}(\alpha)R_{y}(\beta)R_{x}(\gamma) = 
\left[
\begin{array}{ccc}
   cos\alpha & -sin\alpha & 0\\
   sin\alpha &  cos\alpha & 0\\
   0 & 0 & 1 \\
\end{array}
\right] 
\left[
\begin{array}{ccc}
   cos\beta & 0 & sin\beta\\
   0 & 1 & 0 \\
   -sin\beta & 0 & cos\beta \\
\end{array}
\right] 
\left[
\begin{array}{ccc}
   1 & 0 & 0 \\
   0 & cos\gamma & -sin\gamma \\
   0 & sin\gamma &  cos\gamma \\
\end{array}
\right] 
\end{equation}
where $\alpha, \beta$, and $\gamma$ are rotation angles from each axis. We set $\gamma$ to 0 to prevent multiple calculations for one specific angle combination.

For the 4-dimensional rotation matrix, fundamental matrices are for each plane in 4 dimensions. Thus six fundamental matrices are presented. The rotation matrix in 4 dimensions is the dot product of six fundamental matrices.

\begin{equation}
\begin{array}{cl}
R=R_{xy}(\alpha)R_{xz}(\beta)R_{xw}(\gamma)R_{yz}(\delta)R_{yw}(\epsilon)R_{zw}(\zeta)  \\
=
\left[
\begin{array}{cccc}
   cos\alpha & -sin\alpha & 0 & 0 \\
   sin\alpha & cos\alpha & 0 & 0 \\
   0 & 0 & 1 & 0 \\
   0 & 0 & 0 & 1 \\
\end{array}
\right] 
\left[
\begin{array}{cccc}
   cos\beta & 0 & -sin\beta & 0 \\
      0 & 1 & 0 & 0 \\
   sin\beta & 0 & cos\beta & 0  \\
   0 & 0 & 0 & 1 \\
\end{array}
\right] 
\left[
\begin{array}{cccc}
   cos\gamma & 0 & 0 & -sin\gamma \\
    0 & 1 & 0 & 0 \\
    0 & 0 & 1 & 0 \\
   sin\gamma & 0 & 0 & cos\gamma  \\
\end{array}
\right]\\
\left[
\begin{array}{cccc}
    1 & 0 & 0 & 0 \\
   0 & cos\delta & -sin\delta & 0 \\
   0 & sin\delta & cos\delta & 0  \\
    0 & 0 & 0 & 1 \\
\end{array}
\right] 
\left[
\begin{array}{cccc}
    1 & 0 & 0 & 0 \\
   0 & cos\epsilon & 0 & -sin\epsilon \\
       0 & 0 & 1 & 0 \\
   0 & sin\epsilon & 0 & cos\epsilon \\
\end{array}
\right] 
\left[
\begin{array}{cccc}
    1 & 0 & 0 & 0 \\
    0 & 1 & 0 & 0 \\
   0 & 0 & cos\zeta & -sin\zeta \\
   0 & 0 & sin\zeta & cos\zeta \\
\end{array}
\right] 
\end{array}
\end{equation}\\

where $\alpha, \beta, \gamma, \delta, \epsilon$ and $\zeta$ are rotation angles from each plane. We used the Python function \texttt{numpy.matmul} for the dot product of the fundamental matrices. Idealized cases (for example, 45 degrees for every axis) were tested to measure the error from the calculating process, and the error was negligibly small ($10^{-5}$).
This rotation matrix set the $\epsilon$ and $\zeta$ as 0 degrees to reduce the calculation time caused by multiple calculations for one combination of rotation angles.

\section{All recipes} \label{sec:allrecip}
Here, we exhibit 2-dimensional distribution and histograms on the $x$-axis with success rate lines for all the best recipes. The 2-dimensional distributions show 2,3 and 4 indicators (vertical direction) in entire, recent, and ancient merger samples (horizontal direction) with observation data points. Specific information about each recipe is available in Table \ref{tab:bestrecipes} of Section \ref{sec:best}.

\begin{figure*}[ht]
\centering
%\begin{tabular}{cc}
\includegraphics[scale=0.6]{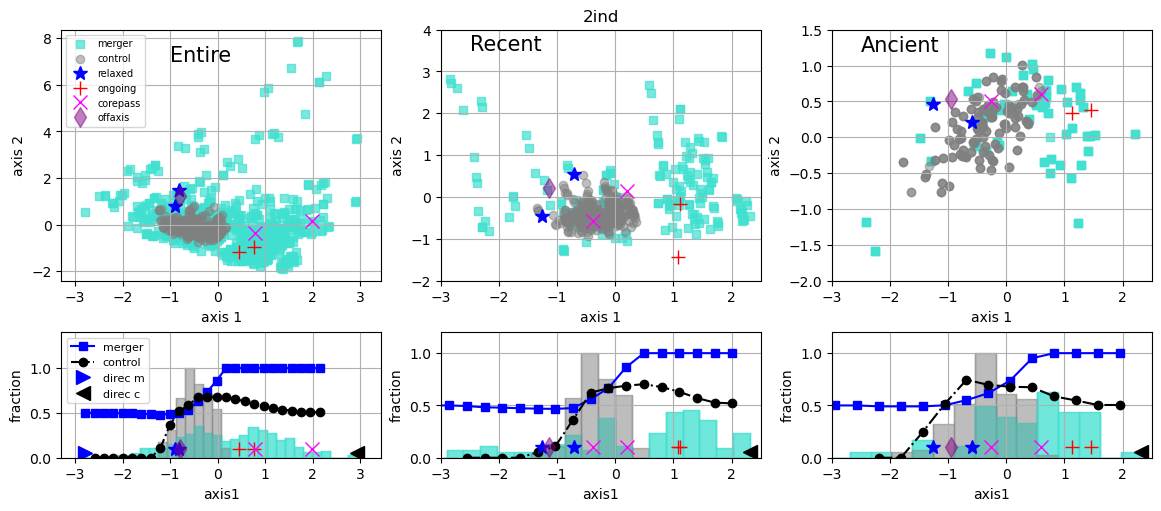}
%\end{tabular}
\caption{ Two-dimensional distribution and histograms with success rate lines of all recipes in 2 indicator combinations (vertical direction) at entire, recent, and ancient merger samples (horizontal direction). Each symbol's information is available in the legend. Detailed interpretation can be seen in Section \ref{sec:best}. }
\end{figure*}\label{fig:allrecip1}
\begin{figure*}[ht]
\centering
%\begin{tabular}{cc}
\includegraphics[scale=0.7]{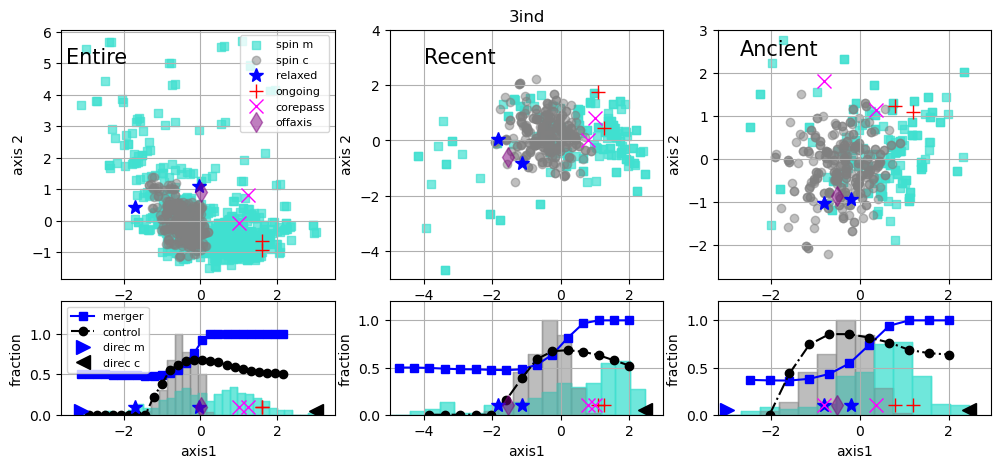}
%\end{tabular}
\caption{ Same with figure \ref{fig:allrecip1}, but using the 3 indicator combinations. }
\end{figure*}\label{fig:allrecip2}
\begin{figure*}[ht]
\centering
%\begin{tabular}{cc}
\includegraphics[scale=0.7]{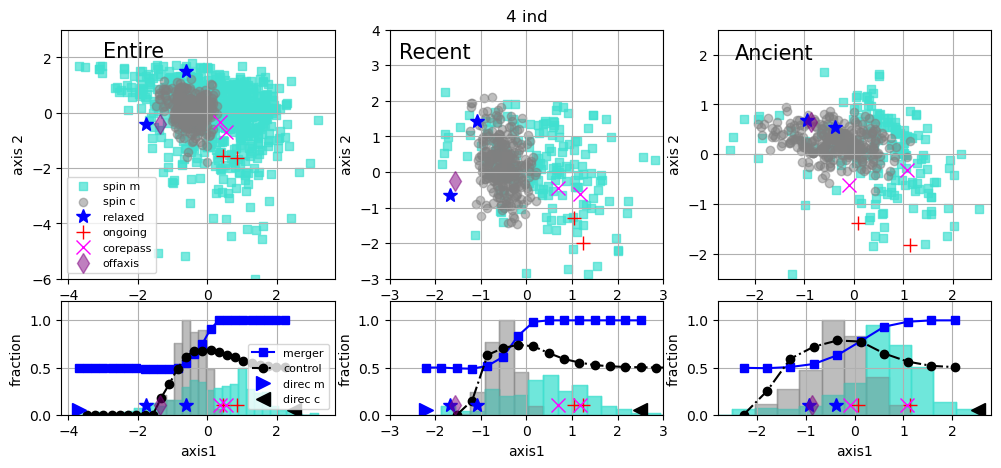}\\
%\end{tabular}
\caption{ Same with figure \ref{fig:allrecip1}, but using the 4 indicator combinations.  }
\end{figure*}\label{fig:allrecip3}

\newpage
\bibliography{new_recipes}{}
\bibliographystyle{aasjournal}



\end{document}